\documentclass[preprint2]{aastex63}

\usepackage{graphicx,epsfig,fancyhdr,psfig,rotating,amsmath,epsf,txfonts,natbib,epstopdf,multirow,booktabs,longtable,color}

\accepted{February 25, 2022}
\submitjournal{ApJ}

\shorttitle{Global EUV Wave}
\shortauthors{Hou et al.}

\def \kms {{\rm km\;s$^{-1}$}}
\def \arcsec {$^{''}$}
\def \ha {H$\alpha$}
\def \heii {He\,{\sc ii}}
\def \sixi {Si\,{\sc xi}}

\graphicspath{{./}}

\begin{document}
\title{Three-dimensional Propagation of the Global EUV Wave associated with a solar eruption on 2021 October 28}

\correspondingauthor{Hui Tian}
\email{huitian@pku.edu.cn}
\correspondingauthor{Jing-Song Wang}
\email{wangjs@cma.gov.cn}
\correspondingauthor{Xiaoxin Zhang}
\email{xxzhang@cma.gov.cn}

\author{Zhenyong Hou}
\affiliation{Key Laboratory of Space Weather, National Center for Space Weather, China Meteorological Administration, Beijing 100081, China}
\affiliation{School of Earth and Space Sciences, Peking University, Beijing, 100871, China}

\author{Hui Tian}
\affiliation{School of Earth and Space Sciences, Peking University, Beijing, 100871, China}
\affiliation{National Astronomical Observatories, Chinese Academy of Sciences, Beijing, 100012, China}

\author{Jing-Song Wang}
\affiliation{Key Laboratory of Space Weather, National Center for Space Weather, China Meteorological Administration, Beijing 100081, China}

\author{Xiaoxin Zhang}
\affiliation{Key Laboratory of Space Weather, National Center for Space Weather, China Meteorological Administration, Beijing 100081, China}

\author{Qiao Song}
\affiliation{Key Laboratory of Space Weather, National Center for Space Weather, China Meteorological Administration, Beijing 100081, China}

\author{Ruisheng Zheng}
\affiliation{School of Space Science and Physics, Institute of Space Sciences, Shandong University, Weihai, Shandong, 264209, China}

\author{Hechao Chen}
\affiliation{School of Earth and Space Sciences, Peking University, Beijing, 100871, China}

\author{Bo Chen}
\affiliation{Changchun Institute of Optics, Fine Mechanics and Physics, Chinese Academy of Sciences, Changchun 130033, China}

\author{Xianyong Bai}
\affiliation{National Astronomical Observatories, Chinese Academy of Sciences, Beijing, 100012, China}

\author{Yajie Chen}
\affiliation{School of Earth and Space Sciences, Peking University, Beijing, 100871, China}

\author{Lingping He}
\affiliation{Changchun Institute of Optics, Fine Mechanics and Physics, Chinese Academy of Sciences, Changchun 130033, China}

\author{Kefei Song}
\affiliation{Changchun Institute of Optics, Fine Mechanics and Physics, Chinese Academy of Sciences, Changchun 130033, China}

\author{Peng Zhang}
\affiliation{Innovation Center for Fengyun Meteorological Satellite, National Satellite Meteorological Center, China Meteorological Administration, Beijing 100081, China}

\author{Xiuqing Hu}
\affiliation{Innovation Center for Fengyun Meteorological Satellite, National Satellite Meteorological Center, China Meteorological Administration, Beijing 100081, China}

\author{Jinping Dun}
\affiliation{Key Laboratory of Space Weather, National Center for Space Weather, China Meteorological Administration, Beijing 100081, China}

\author{Weiguo Zong}
\affiliation{Key Laboratory of Space Weather, National Center for Space Weather, China Meteorological Administration, Beijing 100081, China}

\author{Yongliang Song}
\affiliation{National Astronomical Observatories, Chinese Academy of Sciences, Beijing, 100012, China}

\author{Yu Xu}
\affiliation{School of Earth and Space Sciences, Peking University, Beijing, 100871, China}

\author{Guangyu Tan}
\affiliation{School of Earth and Space Sciences, Peking University, Beijing, 100871, China}

\begin{abstract}
We present a case study for the global extreme ultraviolet (EUV) wave and its chromospheric counterpart `Moreton-Ramsey wave' associated with the second X-class flare in Solar Cycle 25 and a halo coronal mass ejection (CME). The EUV wave was observed in the \ha\ and EUV passbands with different characteristic temperatures.
In the 171\,\AA~and 193/195\,\AA~images, the wave propagates circularly with an initial velocity of 600--720\,\kms\ and a deceleration of 110--320\,m\,s$^{-2}$.
The local coronal plasma is heated from log(T/K)$\approx$5.9 to log(T/K)$\approx$6.2 during the passage of the wavefront.
The \ha\ and 304\,\AA~images also reveal signatures of wave propagation with a velocity of 310--540\,\kms.
With multi-wavelength and dual-perspective observations, we found that the wavefront likely propagates forwardly inclined to the solar surface with a tilt angle of $\sim$53.2$^{\circ}$.
Our results suggest that this EUV wave is a fast-mode magnetohydrodynamic wave or shock driven by the expansion of the associated CME,
  whose wavefront is likely a dome-shaped structure that could impact the upper chromosphere, transition region and corona.
\end{abstract}

\section{Introduction}
\label{sec:intro}
Over the past two decades, coronal wave-like phenomena (i.e., extreme ultraviolet waves, `EUV waves' for short) have been extensively studied \citep{2014SoPh..289.3233L,2016GMS...216..381C}. These EUV waves usually appear as propagating bright or dark fronts in coronal EUV images \citep[e.g.,][]{1999SoPh..190..467W,1999ApJ...517L.151T,2009ApJS..183..225T,2010ApJ...723L..53L,2011ApJ...732L..20C,2012ApJ...748..106T,2012ApJ...752L..23S,2014ApJ...795..130S,2018ApJ...860L...8S,2018ApJ...856...24Y}.
They are sometimes simultaneously observed with Moreton-Ramsey waves \citep{2001ApJ...560L.105W,2002PASJ...54..481E,2004ApJ...608.1124O,2012ApJ...745L..18A,2012ApJ...752L..23S,2019ApJ...873...22S,2019ApJ...882...90L}, which are chromospheric waves observed in \ha\ images \citep{1960AJ.....65U.494M,1960PASP...72..357M,1971ASSL...27..156S,2011fmt..book.....S}.
The EUV waves are often associated with large-scale energetic eruptions, e.g., coronal mass ejections (CMEs) and flares \citep{2002ApJ...569.1009B,2013ApJ...776...58N,2018ApJ...864L..24L}. However, small-scale events such as surges, jets, mini-filaments, and mini-CMEs may also drive EUV waves \citep[e.g.,][]{2011ApJ...739L..39Z,2012ApJ...747...67Z,2012A&A...541A..49Z,2012ApJ...753L..29Z,2012ApJ...753..112Z,2013ApJ...764...70Z,2013MNRAS.431.1359Z,2017ApJ...851..101S}.

The large-scale EUV waves have often been interpreted as fast-mode MHD waves \citep{1999ApJ...517L.151T,2000ApJ...543L..89W,2001ApJ...560L.105W,2002ApJ...574..440O,2012ApJ...746...13L,2018ApJ...864L..24L} driven by CME expansion \citep{2012ApJ...745L..18A,2012ApJ...753...52L,2012ApJ...754....7S,2013ApJ...773L..33S,2018ApJ...860L...8S} and often followed by non-wave components due to the CME-caused magnetic field reconfiguration \citep{2002ApJ...572L..99C,2005ApJ...622.1202C,2012ApJ...752L..23S,2012ApJ...750..134D,2012ApJ...753...52L}.
The EUV waves generally propagate as a single wavefront, while sometimes appear in the form of multiple wavefronts \citep{2012ApJ...753...52L,2019ApJ...873...22S}.
Their propagating velocities are measured to be 50--700\,\kms\ with typical values of 200-400\,\kms\ \citep{2009ApJS..183..225T} from the observations of the Extreme-ultraviolet Imaging Telescope \citep[EIT,][]{1995SoPh..162..291D} onboard the Solar and Heliospheric Observatory \citep[SOHO,][]{1997SoPh..175..571M,1998GeoRL..25.2465T}, and 200--1500\,\kms\ with an averaged value of 644\,\kms\ \citep{2013ApJ...776...58N} from the high-resolution observations with the Atmospheric Imaging Assembly \citep[AIA,][]{2012SoPh..275...17L} onboard the Solar Dynamics Observatory \citep[SDO,][]{2012SoPh..275....3P}.
Previous studies have shown that the EUV waves often experience a deceleration process with a deceleration of an order of several hundred m\,s$^{-2}$ \citep[e.g.,][]{2001ApJ...560L.105W,2002A&A...394..299V,2005ApJ...625L..67V,2008ApJ...681L.113V,2011ApJ...743L..10V,2013ApJ...776...58N}.
As the EUV waves propagate, they could interact with coronal structures. For instance, the wavefronts could not only exhibit reflections\citep{2012ApJ...746...13L,2012ApJ...756..143O,2013ApJ...773L..33S,2013ApJ...775...39Y,2018ApJ...864L..24L,2021arXiv211215098Z}, transmissions\citep{2012ApJ...756..143O,2012ApJ...753...52L,2013ApJ...773L..33S,2018ApJ...864L..24L}, and refractions\citep{2000SoPh..193..161T,2012ApJ...754....7S,2013ApJ...773L..33S}, but also result in oscillations of coronal loops \citep{2012ApJ...753...52L,2012ApJ...754....7S,2013MNRAS.431.1359Z} and filaments \citep{2004ApJ...608.1124O,2012ApJ...745L..18A,2014ApJ...786..151S,2014ApJ...795..130S,2017ApJ...851..101S}. 
EUV waves are often accompanied by quasi-periodic fast-mode propagating wave trains \citep{2014SoPh..289.3233L,2021arXiv211214959S,2022arXiv220108982D}.
Signatures of coronal heating have also been found near wave fronts \citep{1999SoPh..190..467W,2012ApJ...750..134D,2018ApJ...864L..24L}.

Limb observations have shown that an EUV wave does have a dome-shaped wavefront \citep{2010ApJ...716L..57V,2012ApJ...746...13L,2012ApJ...753...52L,2012ApJ...747L..21S,2014ApJ...786..151S,2021SoPh..296..169Z}.
In the studies of \citet{2010ApJ...716L..57V}, \citet{2012ApJ...746...13L}, and \citet{2014ApJ...786..151S}, the upward expansions of wavefronts were found to be faster than the lateral expansions.
With AIA observations, \citet{2012ApJ...753...52L} studied an off-limb EUV wave with an initial height of 110\,Mm driven by a dome-shaped CME front and found that the wavefront propagates forwardly inclined towards the surface of the Sun \citep{2018ApJ...864L..24L}.

\begin{figure*}
\centering
\includegraphics[trim=0.0cm 0.3cm 0.0cm 0.0cm,width=1.0\textwidth]{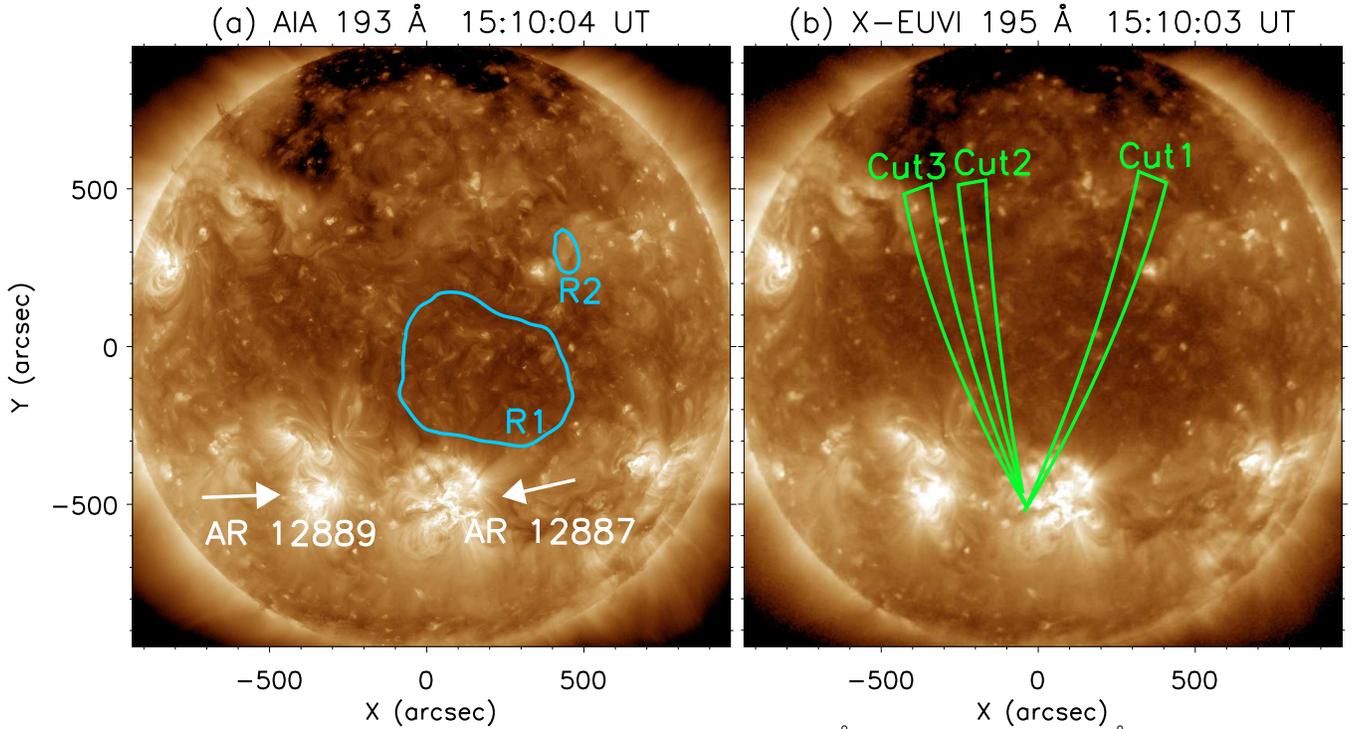}
\caption{Overview of the solar corona before the flare in the AIA 193\,\AA\ (a) and X-EUVI 195\,\AA\ (b) images.
In (a), the white arrows indicate the locations of AR 12887 and AR 12889 and the cyan contours outline two regions used to obtain the normalized intensity variations at the AIA EUV passbands shown in Figure\,\ref{fig:heating}.
In (b), the green lines represent three sector cuts used to obtain the time-distance diagrams shown in Figure\,\ref{fig:stimg}.}
\label{fig:aia193_195}
\end{figure*}

With stereoscopic observations in dual perspectives and multi-passbands,
 here we report one global EUV wave observed at the beginning of Solar Cycle 25, and investigate the three-dimensional (3D) propagation of the wavefront as well as its impact on the coronal plasma temperature.
We describe our observations in Section\,\ref{sec:obs}, present the analysis results in Section\,\ref{sec:res}, 
  discuss the results in Section\,\ref{sec:dis}, and summarize our findings in Section\,\ref{sec:sum}.

\section{Observations}
\label{sec:obs}

\begin{figure*}
\centering
\includegraphics[trim=0.0cm 1.1cm 0.0cm 0.0cm,width=0.97\textwidth]{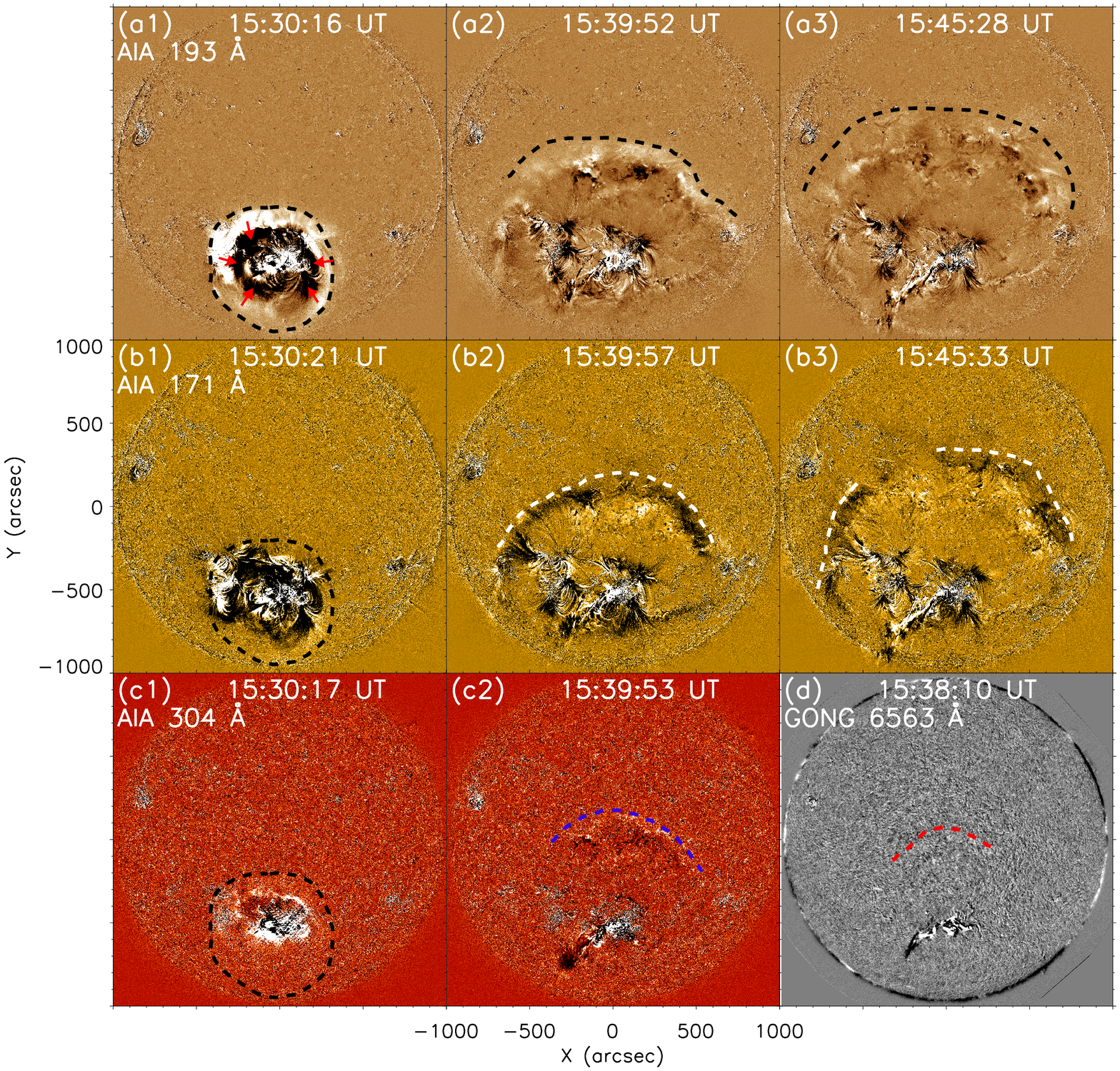}
\caption{Propagation of the EUV wave from the top-down view.
The images are the running-difference images of the AIA 193\,\AA, 171,\AA, 304\,\AA, and GONG \ha\ 6563\,\AA\ passbands.
The black, white, blue, and red dashed lines outline the wavefronts obtained from the AIA 193\,\AA, 171\,\AA, 304\,\AA, and \ha\ running-difference images, respectively.
An animation of this figure is available, showing the propagation of the EUV wave from the top-down view.
It covers a duration of $\sim$43 min from 15:12 UT to 15:55 UT on 2021 October 28.}
\label{fig:aias_ha}
\end{figure*}

\begin{figure*}
\centering
\includegraphics[trim=0.0cm 0.4cm 0.0cm 0.0cm,width=0.98\textwidth]{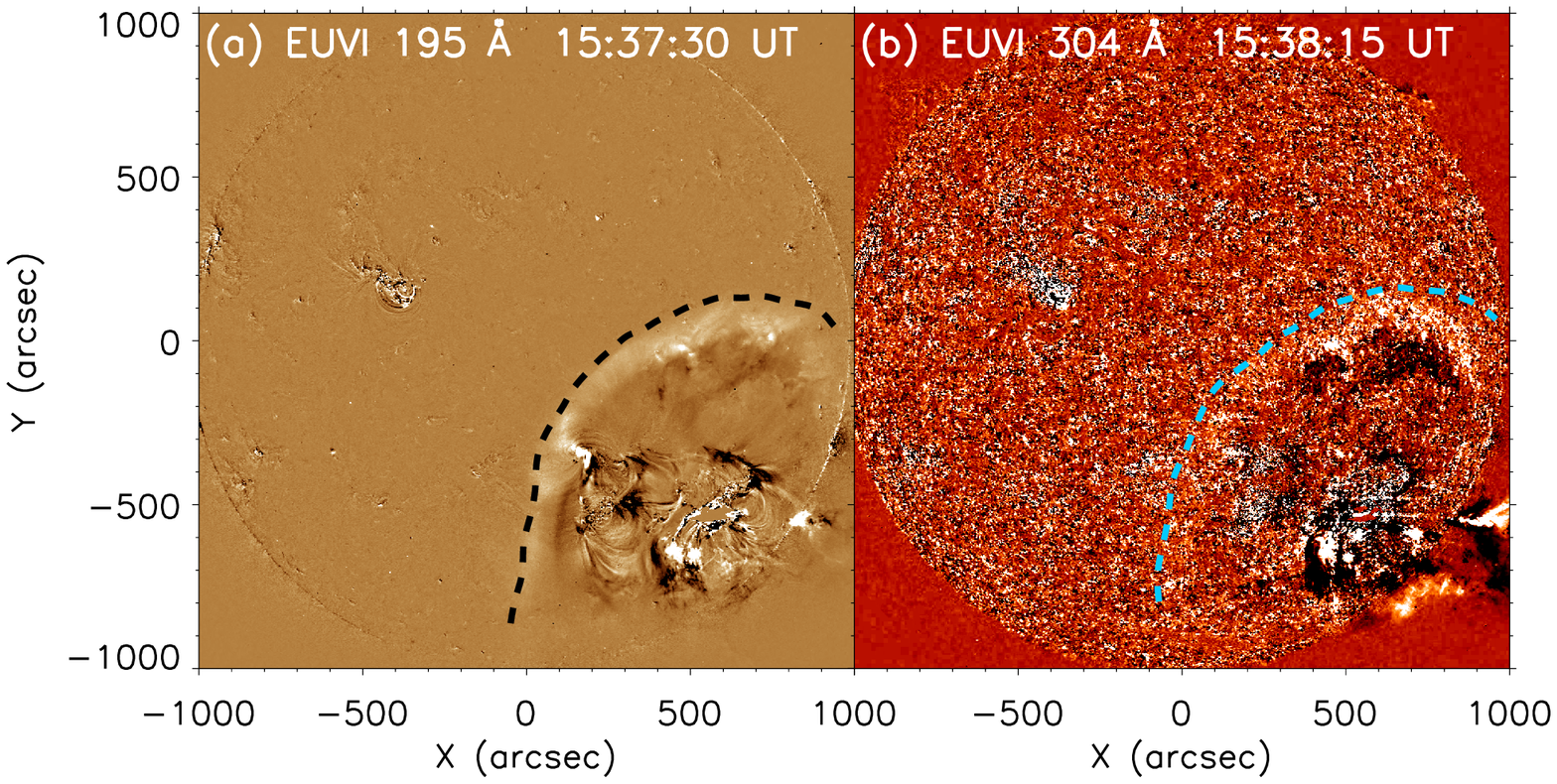}
\caption{Propagation of the EUV wave from the edge-on view.
The black and cyan dashed lines outline the wavefronts obtained from the EUVI 195\,\AA\ (a) and 304\,\AA\ (b) running-difference images.
An animation of this figure is available, showing the propagation of the EUV wave from the edge-on view.
It covers a duration of $\sim$53 min from 15:12 UT to 16:05 UT on 2021 October 28.}
\label{fig:stereo}
\end{figure*}

The global EUV wave of interest occurred on 2021 October 28, 
  was simultaneously observed by AIA onboard SDO,
  the Extreme Ultraviolet Imager \citep[EUVI,][]{2008SSRv..136...67H} onboard the Solar Terrestrial Relations Observatory \citep[STEREO,][]{2008SSRv..136....5K}, 
  the Solar X-ray Extreme Ultraviolet Imager (X-EUVI) onboard the Fengyun-3E saterllite \citep[FY-3E,][]{AAS-2021-0304}, and the Global Oscillation Network Group \citep[GONG,][]{2011SPD....42.1745H}.
From the Earth's view, the full-disk EUV images taken by AIA and X-EUVI revealed the propagations of the EUV wave from the top-down view.
Images of all the AIA EUV passbands, with a 12\,s cadence and a $\sim$0.6\arcsec pixel size, were used in this study.
Different from the AIA 193\,\AA\ passband centered at 193\,\AA,  the EUV image taken by X-EUVI onboard FY-3E in an Early Morning Orbit is centered at 195.5\,\AA\ with a bandwidth of 7.5\,\AA.
It is the only EUV passband of the X-EUVI instrument.
The preprocessing procedure of the X-EUVI 195\,\AA\ image is described in \citet{Song2022a}.
The X-EUVI 195\,\AA\ image has a cadence of $\sim$14\,s and a pixel size of $\sim$2.5\arcsec.
The X-EUVI data will be released to the community after the satellite on-orbit commission test is completed.
We also analyzed the \ha\ images provided by GONG to investigate the chromospheric counterpart of the EUV wave.
The \ha\ images with a cadence of 60\,s and a pixel size of 1.07\arcsec\ were taken with a filter centered at 6562.8\,\AA.

In the meantime, on the western side of the Earth, STEREO-Ahead was separated about 37.5$^{\circ}$ from the Earth-Sun line.
The 195\,\AA\ and 304\,\AA\ images taken by STEREO-Ahead/EUVI revealed the propagation of the EUV wave from the edge-on view.
The pixel size of these images is 1.58\arcsec, and the cadence is $\sim$2.5 min.
 
Since the X-EUVI 195\,\AA\ and the AIA 193\,\AA\ images have similar temperature responses, we aligned them through a linear Pearson correlation.
For aligning the \ha\ images and the AIA images, the sunspot and its surrounding plage region within AR 12887 in the \ha\ and AIA 1600\,\AA\ images were used as the reference features.

\section{Results}
\label{sec:res}

Figure\,\ref{fig:aia193_195} presents an overview of the solar corona at 15:10\,UT in the AIA 193\,\AA\ and X-EUVI 195\,\AA\ passbands, which includes four active regions (ARs), a polar coronal hole, and the rest quiet-Sun regions.
Among them, NOAA AR\,12887 was located in the southern hemisphere and close to AR\,12889.
AR\,12887 was very active around Oct 28, generating a global EUV wave we study here as well as its associated X1.0 flare and halo CME.
The X1.0 flare started and peaked at 15:17\,UT and 15:35\,UT, respectively.
It is the first X-class flare that occurred around the disk center in Solar Cycle 25, and its eruption mechanism is investigated by \citet{Song2022b}.
The global EUV wave appeared at $\sim$15:28\,UT and quickly propagated to the surrounding areas, affecting almost the entire front side of the Sun. From coronagraph images taken by SOHO/LASCO, we can see an associated halo CME that first appeared around 15:53\,UT.

The stereoscopic observations of multiple spacecraft with high spatial and temporal resolutions revealed the evolution of this global EUV wave very well, allowing us to investigate its 3D propagation and its impact on the coronal plasma.

\subsection{Overview of the EUV wave}
\label{subsec:overview}

\begin{figure*}
\centering
\includegraphics[trim=0.0cm 0.8cm 0.0cm 0.0cm,width=0.95\textwidth]{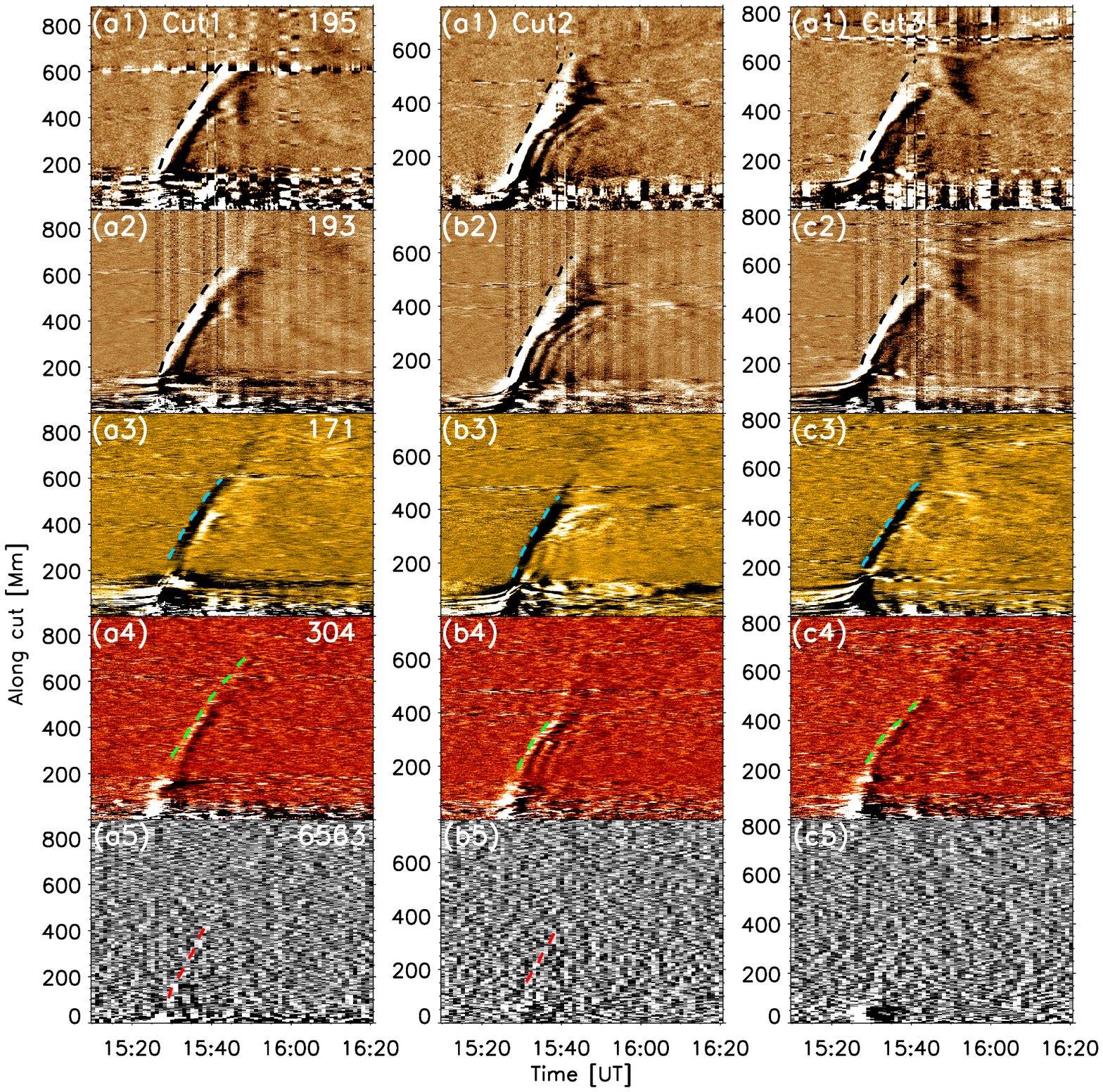}
\caption{Time-distance diagrams showing the propagation properties of the EUV wave, as obtained from the X-EUVI 195\,\AA, AIA 193\,\AA, 171\,\AA, 304\,\AA, and GONG \ha\ 6563\,\AA\ running-difference images.
For the AIA 193\,\AA, 171\,\AA, 304\,\AA, and GONG \ha\ 6563\,\AA\ time-distance diagrams, 
  the dashed lines in different colors represent the tracks of the propagating wavefronts.
The dashed lines overplotted in the X-EUVI 195\,\AA\ time-distance diagrams are the same as those in the AIA 193\,\AA\ time-distance diagrams.
}
\label{fig:stimg}
\end{figure*}

\begin{figure*}
\centering
\includegraphics[trim=0.0cm 0.5cm 0.0cm 0.0cm,width=1.0\textwidth]{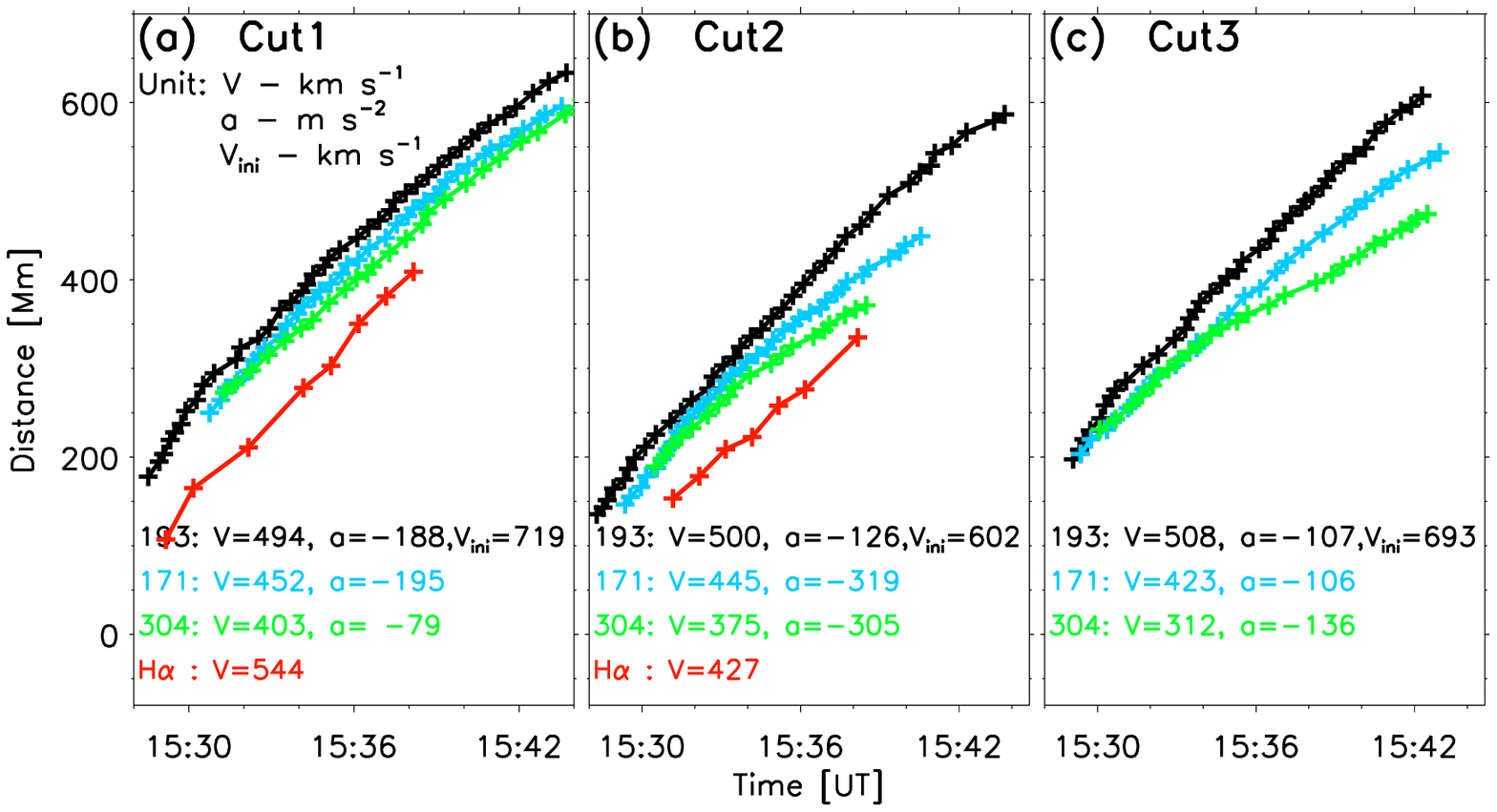}
\caption{Tracks of the propagating wavefronts. 
The tracks in the AIA 193\,\AA, 171\,\AA, 304\,\AA, and GONG \ha\ 6563\,\AA\ passbands are the dotted lines in Figure\,\ref{fig:stimg}.
In each panel, `V' and `V$_{ini}$' indicate the averaged and initial velocities of the EUV wave in \kms\ as derived from linear fits to the whole and starts of tracks, respectively, 
  while `a' indicates the accelerations of the EUV wave in m s$^{-2}$ as derived from quadratic polynomial fits to the whole tracks.}
\label{fig:track}
\end{figure*}

The morphological evolution of the remarkable global EUV wave from the top-down view is shown in Figure\,\ref{fig:aias_ha} and the associated animation.
To highlight the wavefronts, we drew colored dashed lines outlining the outer edge of wavefronts in the AIA 193\,\AA, 171\,\AA, 304\,\AA, and \ha\ running-difference images (2-min seperation).
The colored dashed lines in Figure\,\ref{fig:aias_ha} were visually determined.
The EUV wave first appeared around 15:28\,UT and propagated outwards as a circular wavefront, which was clearly detected in the 193\,\AA\ and 171\,\AA\ passbands.
Meanwhile, a dimming region followed behind the EUV wave, evolving from a small region to a ring-shaped region, as marked by the red arrows in Figure\,\ref{fig:aias_ha}\,(a1).
For a detailed evolution of the EUV wave and the dimming region, we refer to the associated animation of Figure\,\ref{fig:aias_ha}.
The 193\,\AA\ passband reveals the initial stage of the wavefront propagation most clearly (see the black dashed line in Figure\,\ref{fig:aias_ha}\,(a1)), 
  while the 171\,\AA, 304\,\AA, and \ha\ passbands show weak wave signatures.
As the northern part of the EUV wave continues to propagate, its southern part is hard to see.
In addition, as shown in Figure\,\ref{fig:aias_ha} and the associated animation, the wavefront interacts with various coronal structures, resulting in not only the commonly observed reflections/refractions but also emission perturbations such as brightening in 193\,\AA\ and darkening in 171\,\AA.

We also present the morphological evolution of the EUV wave observed by the EUVI 195\,\AA\ and 304\,\AA\ passbands from the edge-on view in Figure\,\ref{fig:stereo} and the associated animation.
The wavefronts in 195\,\AA\ and 304\,\AA\ are also highlighted by the colored dashed lines in Figure\,\ref{fig:stereo}.
The animation reveals that the wavefront appears as a dome-like structure in the EUVI 195\,\AA\ image around 15:30\,UT and evolves globally.
The wavefront in 304\,\AA\ experiences an evolution similar to that in 195\,\AA.

Notably, the northern part of the wavefront is more evident than its southern part 
  in Figure\,\ref{fig:aias_ha} and Figure\,\ref{fig:stereo} (and their associated animations).
Particularlly, only the northern wavefront was observed by the AIA 304\,\AA\ and \ha\ passbands in the form of arc-shaped structures.
Therefore, we focus on the propagation of the northern part of this EUV wave.

\subsection{3D propagation of the EUV wave}
\label{subsec:3D}

\begin{figure*}
\centering
\includegraphics[trim=0.0cm 0.3cm 0.0cm 0.0cm,width=1.0\textwidth]{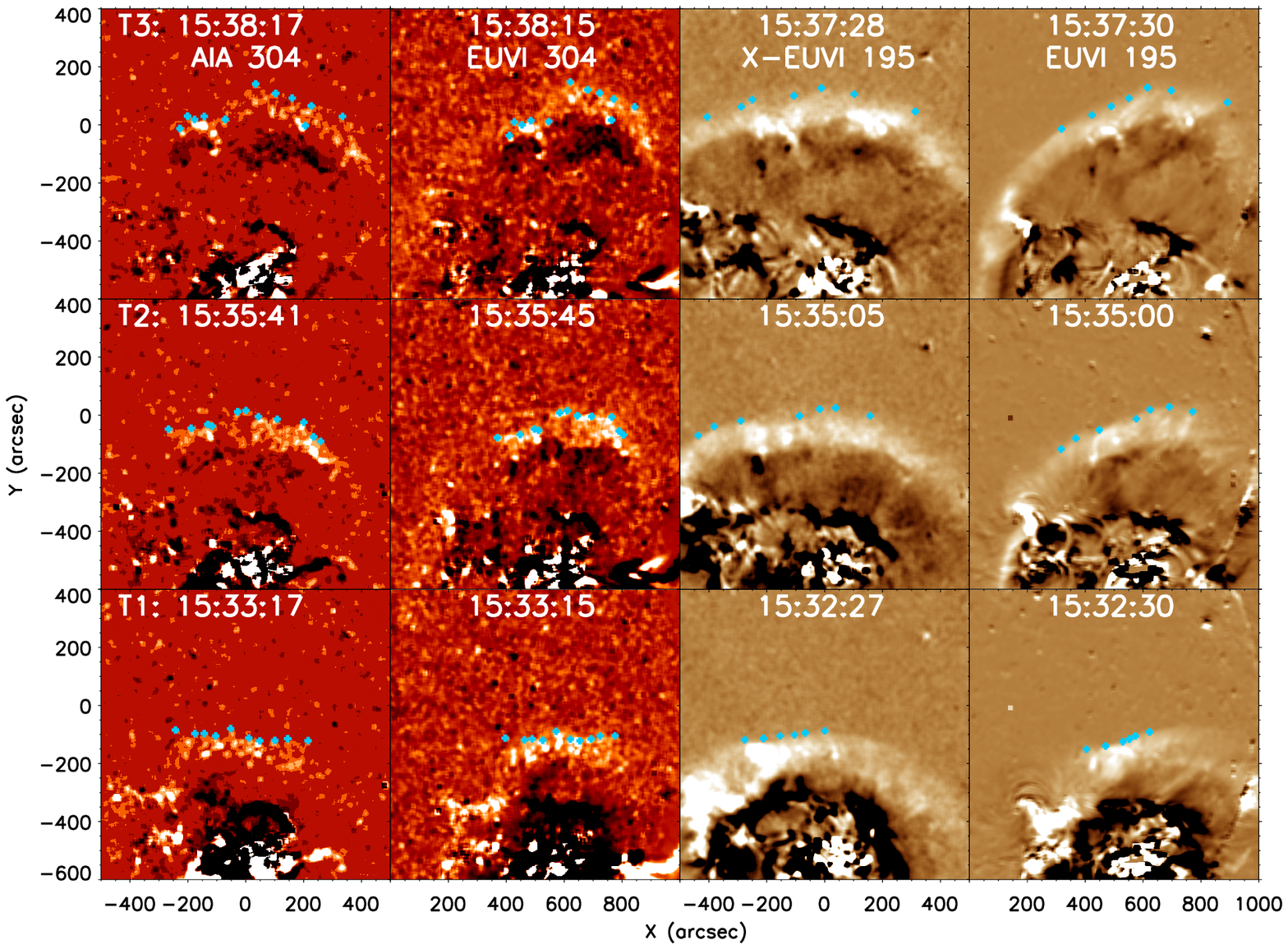}
\caption{Triangulations of the wavefronts in the 304\,\AA\ and 195\,\AA\ passbands.
From bottom to top, three times $\sim$15:33\,UT, $\sim$15:35\,UT, and $\sim$15:38\,UT are defined as T1, T2, and T3.
The cyan symbols are selected to indicate locations of the wavefronts.}
\label{fig:triangulation}
\end{figure*}

We have defined three sector cuts originating from the eruption center and perpendicular to the wavefronts, as indicated by the green triangles in Figure\,\ref{fig:aia193_195}\,(b), to obtain time-distance diagrams.
Figure\,\ref{fig:stimg} shows the time-distance diagrams derived from the running-difference images of the X-EUVI 195\,\AA, AIA 193\,\AA, 171\,\AA, 304\,\AA, and \ha\ passbands,
  which could reflect the propagation of the EUV wave in the plane of the sky from the top-down view.
The distances were estimated along the sector cuts by considering the curvature of the solar surface. 
In the time-distance diagrams, the colored dotted lines mark the propagating tracks of the outer edges of the wavefronts in different images, which were visually determined.

The propagations of the wavefronts are manifested as evident, wide stripes in the X-EUVI 195\,\AA, AIA 193\,\AA, and 171\,\AA\ images and weak, narrow ones in the AIA 304\,\AA\ and \ha\ images.
From Figure\,\ref{fig:stimg} we can see that the propagating signatures appear about 150\,Mm away from the eruption center. The wavefront appears to be bright in 195/193\,\AA\ and dark in 171\,\AA.
As shown in the first and second rows of Figure\,\ref{fig:stimg}, the overplotted black dotted lines were both obtained from the AIA 193\,\AA\ time-distance diagrams, revealing that the propagating tracks in 195\,\AA\ and 193\,\AA\ coincide well with each other.

We further plotted the propagating tracks of the wavefronts along Cut\,1, Cut\,2, and Cut\,3 in Figure\,\ref{fig:track}.
To quantitatively determine the average velocities and accelerations, we applied linear fits and quadratic polynomial fits to the propagating tracks, respectively.
The resulting velocities and accelerations for different sector cuts and passbands are also shown in Figure\,\ref{fig:track}. The propagating tracks in the \ha\ time-distance diagrams are too weak for an accurate calculation of the acceleration.
It can be seen that the EUV wave fronts propagate at velocities of 420--510 \kms\ and decelerations of 110--320 m\,s$^{-2}$ in the AIA 193\,\AA\ and 171\,\AA\ passbands.
We also calculated the propagating velocity of the EUV wave during its initial propagation stage during 15:28--15:32\,UT and obtained a velocity of 600--720\,\kms.
Our results are consistent with the previous studies by \citet{2011A&A...532A.151W} and \citet{2014SoPh..289.4563M}, in which they found that some events with fast initial velocities ($>$300\,\kms) show distinct deceleration (several hundred m\,s$^{-2}$) during their propagations.
On the other hand, the chromospheric counterpart of the EUV wave propagates in the \ha\ passband at a velocity of 310--540 \kms, which is similar to that in the AIA 304\,\AA\ passband.

We also found that the propagating tracks in 193\,\AA\ significantly precede those in the other passbands with lower temperatures such as 304\,\AA\ and \ha (see the colored tracks in Figure\,\ref{fig:track}).
Though not explicitely mentioned in \citet{2012ApJ...745L..18A}, we noticed a similar behavior in their event, in which the EUV wavefront slightly preceded the \ha\ wavefront.
This indicates that the wavefronts with different temperatures propagate to different distances within the same period of time, i.e., the wavefront with a higher temperature propagated ahead of those with lower temperatures.
From Figure\,\ref{fig:track} we calculated the differences of the plane-of-sky distances in the 193\,\AA\ and 304\,\AA\ passbands for Cuts\,1--3 in the time periods of 15:32--15:38 UT, which are 43.9$\pm$7.5\,Mm, 55.2$\pm$21.1\,Mm, and 51.5$\pm$23.4\,Mm, respectively.
We take the averaged value 50.5\,Mm for further analysis.

As mentioned above, the EUV wave was observed in dual perspectives, allowing us to derive the height difference
  between the wavefronts with different temperatures (e.g., 195\,\AA\ from FY-3E/X-EUVI and STEREO-Ahead/EUVI; 304\,\AA\ from SDO/AIA and STEREO-Ahead/EUVI).
Using the stereoscopic observations, we reconstructed the 3D structures of the wavefronts in the 195\,\AA\ and 304\,\AA\ passbands, and derived the heights of the wavefronts in these two passbands
  through the \textit{SSW} procedure `scc\_measure.pro'.
As shown in Figure\,\ref{fig:triangulation}, we measured the wavefront heights at three different times in the 195\,\AA\ and 304\,\AA\ passbands by choosing several points (marked by the cyan plus symbols) along the outer edges of the wavefronts.
The derived heights of the wavefronts in 304\,\AA\ are 3.5$\pm$1.3\,Mm (T1), 3.6$\pm$2.0\,Mm (T2), and 4.2$\pm$1.7\,Mm (T3),
  while the heights in 195\,\AA\ have been estimated to be 37.2$\pm$17.9\,Mm (T1), 89.3$\pm$44.0\,Mm (T2), and 87.0$\pm$41.2\,Mm (T3).
The wavefront heights in 195\,\AA\ are consistent with those reported by \citet{2009SoPh..259...49P}, \citet{2009ApJ...703L.118K}, \citet{2012ApJ...753...52L}, and \citet{2012ApJ...754....7S}.
Based on these results, we calculated the average height differences of the wavefronts between the 195\,\AA\ and 304\,\AA\ passbands, which are 33.8\,Mm, 85.7\,Mm, and 82.8\,Mm at T1, T2, and T3, respectively. We take the average value 67.4\,Mm as the distance between the wavefronts in the two passbands in the radial direction.

This, combined with the difference in the plane-of-sky distances of the wavefronts between the two passbands,
  indicates that the wavefront propagated forwardly inclined to the solar surface with an averaged tilt angle of arctan(67.4/50.5)=53.2$^{\circ}$ in the time period of 15:32--15:38 UT.

\subsection{Heating of the coronal plasma by the EUV wave}
\label{subsec:heating}

\begin{figure*}
\centering
\includegraphics[trim=0.0cm 0.7cm 0.0cm 0.0cm,width=1.0\textwidth]{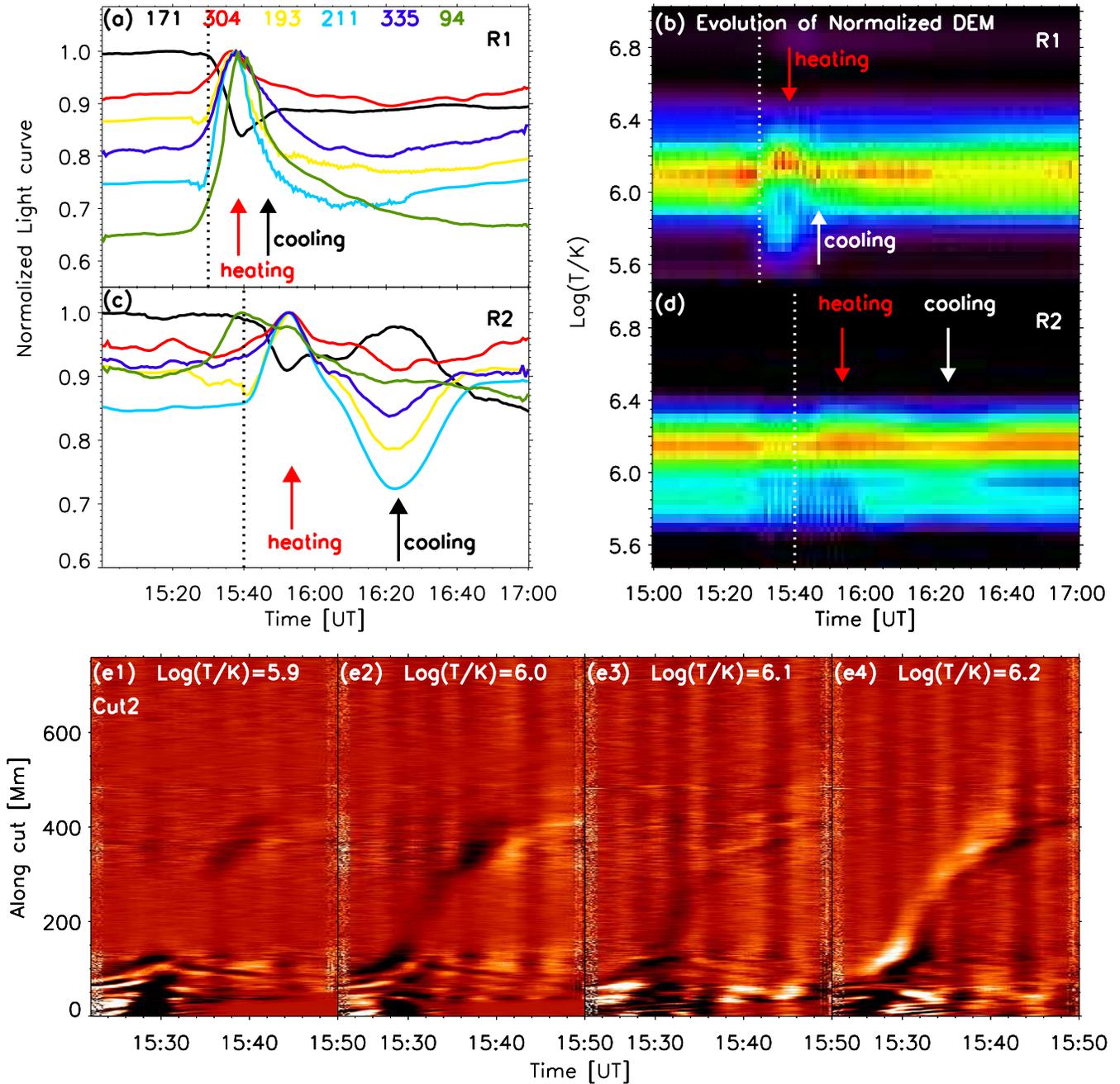}
\caption{Temperature change of the coronal plasma caused by the EUV wave.
Panels (a) and (c) show the normalized intensity variations in the AIA EUV passbands for regions `R1' and `R2', respectively.
Panels (b) and (d) present the temporal evolution of the DEM for regions `R1' and `R2', respectively.
The red and black arrows indicate the plasma heating and cooling phases, respectively.
The vertical dotted lines mark the times of EUV wave passage.
Panels (e1)--(e4) show the time-distance diagrams along Cut\,2 from the running-difference EM maps at four temperatures (i.e., log(T/K)=5.9, 6.0, 6.1, and 6.2).}
\label{fig:heating}
\end{figure*}

In Section\,\ref{subsec:overview} and Section\,\ref{subsec:3D}, we detected an opposite intensity change
  between the 193/195\,\AA\ and 171\,\AA\ passbands, as shown in Figure\,\ref{fig:aias_ha} and Figure\,\ref{fig:stimg}.
We have analyzed the thermal response of the corona plasma to the EUV wave in detail, and shown the results in Figure\,\ref{fig:heating}.

The top and middle rows of Figure\,\ref{fig:heating} present normalized light curves in different AIA passbands for a quiet-Sun region  (`R1') and a loop footpoint region (`R2') marked by the cyan contours in Figure\,\ref{fig:aia193_195}\,(a). These two regions were impacted by the EUV wave.
When the EUV wave reaches `R1' around 15:30\,UT, the intensities in 193\,\AA, 211\,\AA, 335\,\AA, and 94\,\AA\ all increase substantially, while the 171\,\AA\ intensity decreases.
After 15:40\,UT, the 193/211/335/94\,\AA\ intensities begin to decrease and the 171\,\AA\ intensity increases.
Since the temperature responses of the 193\,\AA\ and 211\,\AA\ passbands peak at higher temperatures compared to the 171\,\AA\ passband, this result likely suggests a plasma heating and cooling cycle in the corona \citep[e.g.,][]{1999SoPh..190..467W,2012ApJ...746...13L,2012ApJ...750..134D,2012ApJ...753...52L,2018ApJ...864L..24L}, in which the plasma in the corona experiences heating from log(T/K)$\approx$5.9 to log(T/K)$\approx$6.2 before cooling.
The 304\,\AA\ intensity also increases as the EUV wave reaches, which might be related to the compression of the transition region plasma.

To examine the emission changes at different temperatures caused by the EUV wave,
  we also performed a differential emission measure (DEM) analysis \citep{2015ApJ...807..143C,2018ApJ...856L..17S,2020ApJ...898...88X,2021Innov...200083S} for `R1'. The result is shown in Figure\,\ref{fig:heating}\,(b). We can see that the DEM around log(T/K)$\approx$6.2 increases after 15:30 UT and decreases after 15:40 UT. Simultaneously, the DEM around log(T/K)$\approx$5.9 shows an opposite change, indicating the plasma heating and subsequent cooling process. When the EUV wave reaches `R2' around 15:40\,UT, we see similar results, as shown in Figure\,\ref{fig:heating}\,(c) and (d). 
This means that the local plasma at the loop footpoint region `R2' may have also been heated from log(T/K)$\approx$5.9 to log(T/K)$\approx$6.2 by the EUV wave.

We further present the time-distance diagrams along Cut\,2 from the running-difference EM maps at four temperatures (log(T/K)=5.9, 6.0, 6.1, and 6.2) in Figure\,\ref{fig:heating}\,(e1)--(e4), respectively.
We see a dark propagating track of the wavefront at the temperatures of log(T/K)=5.9\&6.0 and a bright propagating track at the temperature of log(T/K)=6.2, indicating plasma heating from log(T/K)$\approx$5.9 to log(T/K)$\approx$6.2.

To determine whether the heating and cooling processes associated with the observed wavefronts are adiabatic, we examined the adiabatic relation, i.e., $T\rho^{\gamma-1}$ =Const, which gives $\delta T/T=(\gamma-1)\delta \rho/\rho$. Here $\delta T/T$ and $\delta \rho/\rho$ represent the relative changes of temperature and density, respectively. The adiabatic index $\gamma$ is 5/3.
Through a DEM analysis, \citet{2021ApJ...911..118D} found that perturbations of the density and temperature in an observed wavefront are 24\% and 15\%, respectively, consistent with an adiabatic process.
Here we took the wavefront height ($\sim$87\,Mm at 15:38\,UT) in 195 Å as a rough estimation of the line-of-sight integration length, and calculated the density in region R2.
We also estimated the temperature and its variation by integrating the DEM over the temperature range of log(T/K)=5.5--6.8.
The variations of the density and temperature after the passage of the wavefront were estimated to be 3\%--4.5\% and 2\%--2.8\%, respectively, which are also consistent with an adiabatic process. However, this conclusion may need to be taken with caution as the density and temperature estimations are likely subject to large uncertainties. 

\section{Discussion}
\label{sec:dis}

In the last twenty years, the EUV waves have been extensively studied. Despite intensive debates on the nature of the EUV waves, in recent years more and more researchers tend to interpret these waves by a `hybrid model', in which an outer wave component is likely a fast-mode MHD wave driven by the CME expansion and an inner non-wave component is caused by the magnetic field reconfiguration during the CME eruption \citep{2012ApJ...750..134D,2012ApJ...753...52L,2014SoPh..289.3233L,2016GMS...216..381C}.
The primary EUV wave we studied is accompanied by an X1.0 flare and a halo CME.
It originates from a distance of $\sim$150\,Mm away from the eruption center, where the accompanying dimming region stops expanding and the wavefront continues to propagate with an initial velocity of 600--720\,\kms\ and a significant deceleration of 110--320\,m\,s$^{-2}$.
The initial velocity is larger than the local sound speed ($\sim$\,150\,\kms), and possibly also larger than the coronal Alfv\'en speed \citep[200--400\,\kms,][]{2012ApJ...744...72G}.
These indicate that the EUV wave is likely a nonlinear fast-mode MHD wave or shock driven by the CME expansion \citep{2012ApJ...745L..18A,2012ApJ...753...52L,2012ApJ...754....7S,2013ApJ...773L..33S,2018ApJ...860L...8S}.
Assuming that this EUV wave decays to an ordinary fast-mode wave, its final velocity can be taken as a proxy for the fast-mode speed.
Then, the fast-mode Mach number is estimated to be 1.52--1.82 for this event by dividing the initial velocity by the final one (estimated to be 380--440\,\kms).

The EUV wave is likely a 3D structure propagating along the solar surface with a fast-mode speed.
The wavefronts observed in the 193\,\AA\ and 171\,\AA\ passbands are the line-of-sight integrations projected on the solar disk at the temperature of 1.5 MK and 1 MK, respectively.
As reported by \citet{2005A&A...435.1123W} and \citet{2021ApJ...911..118D}, the fast-mode speed increases with height and leads to the wavefront at larger heights having a faster apparent motion on the solar surface.
Our result that the wavefront seen at 193\,\AA\ propagates faster than the one in 171\,\AA\ is consistent with this when we consider the effect that temperature has on the observability of the wavefronts.
The densities of plasma at various temperatures show a different fall-off with height due to the different hydrostatic scale height.
Thus, emission of the cooler plasma at 171\,\AA\ is more concentrated at lower heights than emission at 193\,\AA\, and the same holds for the observed wavefronts.
Therefore, the different propagation speeds determined from the locations of the wavefronts in 193\,\AA\ and 171\,\AA\ might well be a result of this hydrostatic weighting bias.
For 304\,\AA, the presence of emission lines with different formation temperatures in the 304\,\AA\ passband makes it hard to determine the dominated emission for the wave signatures.
However, we found that the wavefront has a chromospheric height and is preceded by and lower than the 193\,\AA\ wavefront, indicating that the 304\,\AA\ wavefront is likely dominated by the \heii\ line.
For the chromospheric counterpart Moreton-Ramsey wave, the wavefront signature is dominated by the chromospheric \ha\ line and also clearly preceded by the 193\,\AA\ wavefront. 
Then, the difference of the propagating speeds between the coronal and chromospheric/transition region wavefronts would be also an effect of temperature.

\citet{2012ApJ...752L..23S} reported an EUV wave simultaneously observed in the 1600/1700\,\AA, \ha, 193\,\AA, and 211\,\AA\ filters,
  covering a broad temperature range from the photosphere to the low corona.
They found that the EUV wave had similar initial propagating velocities in the photospheric and coronal passbands, 
  and suggested that the coronal wave and its photospheric/chromospheric counterparts had a common origin.
We also observed a chromospheric counterpart of the EUV wave in the \ha\ images, indicating that the EUV wave is accompanied by a Moreton-Ramsey wave.
However, no wave signature appeared in the 1600/1700\,\AA\ images from SDO/AIA.
This implies that the EUV wave in our study may have affected the upper chromosphere but not deeper in the atmosphere.

Several previous observations have shown that EUV waves could be detected in the 304\,\AA\ passband \citep{2010ApJ...723L..53L,2011ApJ...741L..21L,2012ApJ...752L..23S,2012ApJ...754....7S,2018ApJ...856...24Y,2019ApJ...882...90L}.
In quiet-Sun regions and ARs, the AIA 304\,\AA\ passband is dominated by two \heii\ 303.8\,\AA\ lines, which are formed in the transition region at a temperature of log(T/K)$\approx$4.7,
  but a coronal line of \sixi\ 303.33\,\AA\ at the temperature of log(T/K)$\approx$6.2 could also make an important contribution to the 304\,\AA\ passband in some cases \citep{2010A&A...521A..21O}.
\citet{2012ApJ...752L..23S} found that the propagation of an EUV wave in the 304\,\AA\ images was similar to that in the 193\,\AA\ passband 
  and suggested that the wavefront in the 304\,\AA\ passband was the manifestation of the contribution of the coronal \sixi\ line rather than the transition region \heii\ line.
However, in our event, we found that the 304\,\AA\ wavefront with a chromospheric height is preceded by and lower than the 193/195\,\AA\ wavefront, suggesting that the 304\,\AA\ wavefront is likely dominated by the \heii\ line.
Also, the 304\,\AA\ wavefront clearly precedes the \ha\ wavefront, suggesting that the 304\,\AA\ passband may be used to reveal the link between the coronal EUV wave and the chromospheric Moreton-Ramsey wave.

For the EUV waves, dome-like wavefronts have been reported by some authors \citep{2010ApJ...716L..57V,2012ApJ...746...13L,2012ApJ...753...52L,2012ApJ...747L..21S,2014ApJ...786..151S,2021SoPh..296..169Z}.
\citet{2010ApJ...716L..57V}, \citet{2012ApJ...746...13L}, and \citet{2014ApJ...786..151S} found that the upward expansions of their studied wavefronts were faster than the lateral expansions of the wavefronts.
With AIA observations, \citet{2012ApJ...753...52L} analyzed an off-limb EUV wave driven by a dome-shaped CME front, and suggested that its wavefront propagated forwardly inclined to the solar surface \citep{2018ApJ...864L..24L}.
Similarly, in our event, the propagating tracks in the warm passbands (e.g., 193\,\AA\ and 195\,\AA) precede those in the cool passbands (e.g., 304\,\AA) by $\sim$36.4\,Mm.
Meanwhile, we found the wavefront in the 195\,\AA\ passband $\sim$67.4\,Mm higher than that in the 304\,\AA\ passband from dual-perspective and multi-passband observations.
Based on these findings, we suggest that the propagating wavefront appears to be a dome-like structure, propagating forwardly inclined toward the solar surface with an averaged tilt angle of $\sim$53.2$^{\circ}$ in the time period of 15:32--15:38 UT. This is consistent with the finding of \citet{2012ApJ...753...52L}, in which the slit angle of the wavefront near the flare site appeared to decrease with decreasing hight, ranging from 73$^{\circ}$ at 72\,Mm to 48$^{\circ}$ at the coronal base.
Also, it has been reported that wavefronts of the EUV waves tend to get increasingly tilted during their propagation \citep{2003SoPh..212..121H,2009ApJ...700L.182P,2009ApJ...703L.118K,2012ApJ...753...52L} due to the increase of fast-mode speed with height in the low corona \citep{2005A&A...435.1123W}.

\section{Summary}
\label{sec:sum}

In this study, we have presented analysis results of a remarkable global EUV wave and its chromospheric counterpart `Moreton-Ramsey wave' with stereoscopic observations from SDO/AIA, STEREO-Ahead/EUVI, FY-3E/X-EUVI, and GONG.
The EUV wave is associated with an X1.0 flare and a halo CME, and appears in the \ha\ and EUV passbands with different characteristic temperatures (e.g., 304\,\AA, 171\,\AA, and 193/195\,\AA).

The EUV wave becomes evident at a distance of $\sim$150\,Mm away from the eruption center and propagates circularly outward with an initial velocity of 600--720\,\kms\ and a significant deceleration of 110--320\,m\,s$^{-2}$.
During the propagation of the EUV wave, the wavefront propagates forwardly inclined to the solar surface with a tilt angle of $\sim$53.2$^{\circ}$.
This implies that the propagating wavefront is likely a dome-like structure, which could interact with the solar atmosphere.
The plasma in the low corona is heated from log(T/K)$\approx$5.9 to log(T/K)$\approx$6.2 on the propagation path of the wavefront. 
The 304\,\AA\ and \ha\ passbands also reveal signatures of the wave, indicating that the EUV wave could affect the low atmosphere.
This event is also accompanied by a dimming region, which just follows the wavefront and evolves from a small region to a ring-shaped region stopping at a distance of $<$150\,Mm away from the eruption center.
Our results suggest that the observed EUV wave is likely a fast-mode MHD wave or shock driven by the expansion of its associated CME.

\acknowledgments
This work was supported by National Key R\&D Program of China No. 2021YFA0718600, NSFC grants 41774195, 41931073, 11825301, and 11790304, Ten-thousand Talents Program of Jing-Song Wang, China Postdoctoral Science Foundation (grant numbers 2021M700246 and 2020M680201), and National Postdoctoral Program for Innovative Talents (BX20200013). AIA is instrument onboard the Solar Dynamics Observatory (SDO), a mission for NASA's Living With a Star program.
We thank the SDO/AIA, FY-3E/X-EUVI, STEREO/SECCHI, and GONG for providing data.

\bibliographystyle{aasjournal}
\bibliography{bibliography}

\end{document}